\documentclass[sigconf, nonacm]{acmart}

\setcopyright{acmlicensed}
\copyrightyear{2026}
\acmYear{2026}
\acmConference[KDD-MLF 2026]{KDD 9th Workshop on Machine Learning in Finance}{August 9–13, 2026}{Jeju, Korea}
\acmDOI{}
\acmISBN{}

\usepackage{xcolor}
\usepackage{booktabs}

\AtBeginDocument{%
  }

\begin{document}

\title[Your Spending Needs Attention: Modeling Financial Habits with Transformers]{Your Spending Needs Attention: \\ Modeling Financial Habits with Transformers}

\author{D. T. Braithwaite\textsuperscript{*}, Misael Cavalcanti\textsuperscript{*}, R. Austin McEver\textsuperscript{*}, Hiroto Udagawa, Daniel Silva, \\ Rohan Ramanath, Felipe Meneses, Arissa Yoshida, Evan Wingert, Matheus Ramos, Brian Zanfelice, Aman Gupta}
\affiliation{%
  \institution{Nubank}
  \city{}
  \country{}
}
\email{{daniel.braithwaite, misael.cavalcanti, austin.mcever}@nubank.com.br}
\renewcommand{\shortauthors}{D. T. Braithwaite, Misael Cavalcanti, R. Austin McEver}

\thanks{\textsuperscript{*}These authors contributed equally and are listed alphabetically by last name.}

\begin{abstract}
The industry standard for credit risk modeling relies on manually transforming raw transactions into tabular features, a process that often results in information loss. We propose nuFormer, a Transformer-based model that learns representations directly from raw financial sequences. Each transaction is encoded as a compact sequence of special tokens for structured attributes (amounts, timestamps) alongside tokenized descriptions, reducing token count by approximately 4× compared to text-only approaches and enabling the model to capture long-range spending patterns. To retain static risk factors, we introduce Joint Fusion, which trains the Transformer end-to-end with a tabular network. On a large-scale credit default prediction task, nuFormer achieves a 1.25\% relative AUC improvement. This model was trained on $\sim$500 billion tokens and deployed to 100\% of customers in one of our markets, serving over 100 million users. The architecture has also driven improvements in different applications (such as lending risk and income prediction) as well as in other countries with different languages. Crucially, these gains are achieved solely through enhanced representation learning rather than incorporating new data sources, validating that automated feature discovery is a scalable, operationally efficient alternative to manual feature engineering in financial systems.
\end{abstract}

\maketitle

\section{Introduction}
Credit risk modeling involves estimating the probability that borrowers will default on their obligations, and determines whether hundreds of millions of consumers gain access to financial products. Prediction errors in credit risk models translate to substantial financial losses and misallocated credit. Yet the richest source of predictive signal, detailed transaction histories, remains underutilized. These records capture spending patterns, payment reliability, and financial stress signals over months or years, but most production systems still compress this sequential data into handcrafted tabular features and then train traditional models such as gradient-boosted trees (GBTs). While this pipeline is simple and relatively interpretable, it is labor-intensive and often discards useful information. It also handles text poorly: signals that are hard to encode as categorical variables (e.g., such as free-form descriptions) are frequently dropped altogether.

In many domains, ML has advanced toward learning representations from raw data. One example is convolutional neural networks, which automatically learn features such as edges, textures, and shapes from raw images \citep{krizhevsky2012imagenet, simonyan2014very}. Similar approaches are employed by YouTube \citep{covington2016deep}, Pinterest \citep{pancha2022pinnerformer}, Meta \citep{rangadurai2022nxtpost}, and Alibaba \citep{li2019multi, pi2019practice} to learn user representations from event sequences. Pre-trained representations extend this, training on massive unlabeled datasets with self-supervised objectives producing effective embeddings across vision \citep{awais2023foundational, radford2021learning}, audio \citep{yang2023uniaudio, das2024speechverse, ao2021speecht5, radford2023robust, zhang2023speechgpt}, and language \citep{touvron2023llama, brown2020language, achiam2023gpt}.

Despite this aforementioned progress in representation learning, financial applications have largely remained dependent on hand-crafted features. One recent work by \citep{babaev2022coles} proposes contrastive learning for event sequences (CoLES) with self-supervision for learning embeddings of user transaction sequences. A subsequent work by \citep{skalski2023towards} proposes an autoregressive next-event prediction approach called NPPR. While both CoLES and NPPR facilitate the automatic discovery of features from raw transaction data, this paper addresses three limitations of these studies.

The first limitation of NPPR and CoLES is that they only utilize categorical or numerical features from the event sequence. On the other hand, we hypothesize that there is much value in learning from natural language features, e.g., descriptions. Secondly, both use RNN models, which struggle to capture long-range dependencies compared to Transformers \citep{vaswani2017attention}, a significant limitation for transaction data with seasonal patterns. Finally, the scale of data in these studies (billions of transactions) is orders of magnitude smaller than what we work with (over 100 billion transactions across 100M+ members). This is important because scaling often improves model quality \citep{kaplan2020scaling, hoffmann2022training}. More recently, PRAGMA \citep{ostroukhov2026pragma} uses a bidirectional Transformer and masked pre-training to jointly model heterogeneous banking events and static profile information, reporting offline results across several downstream tasks. Our work studies causal next-token pre-training with downstream tabular fusion and additionally includes online evaluation in production.

An alternative line of work treats sequential data as natural language. In sequential recommendation, Recformer \citep{li2023text} serializes item attributes into text strings, enabling standard language model pre-training and generalization to unseen items. However, this text-only formulation becomes prohibitive for transaction modeling: serializing amounts, timestamps, and descriptions as natural language requires approximately 55 tokens per transaction, limiting a 2048-context model to roughly 37 transactions. Credit risk patterns (e.g., payment behavior before due dates, seasonal spending shifts, gradual financial stress) often span months of history, requiring observation of hundreds of transactions.

In this paper, we propose nuFormer, a transformer-based model for learning representations from financial transaction sequences. Our contributions are: (1) a compact tokenization scheme that encodes transactions as special tokens for structured attributes alongside BPE-tokenized descriptions, reducing per-transaction token count by ~4$\times$ compared to text-only approaches; (2) Joint Fusion, an end-to-end architecture that combines transformer embeddings with tabular features via a modified DCNv2 network with periodic numerical embeddings; (3) deployment at scale to credit default prediction in one of our markets with 1.25\% relative AUC improvement, and 100M+ users served; and (4) analysis of scaling properties across model size, context length, and data volume.
\section{Related Work}
\label{sec:related_work}

\paragraph{Sequential Recommendation}
Sequential recommendation (SR) systems involve modeling a user's behavior from a sequence of item interactions (e.g., purchases, clicks, etc.) and attempting to predict future actions to recommend these to the user. This is closely related to modeling transaction sequences, as items also consist of a rich set of attributes, such as text, numerical, categorical, and images.

One of the earliest applications of transformer models to SR is SASRec \citep{kang2018self}, which utilizes an ID-based representation of items. In this setup, each item is assigned a unique ID, and item embeddings are obtained from an embedding table. Importantly, item attributes are not utilized. A causal transformer is then trained using a generative next-item prediction objective, which is a self-supervised loss. This approach outperformed other state-of-the-art models on standard SR tasks. However, ID-based approaches suffer from the so-called cold-start problem, where it is not possible to generalize to items unseen during training.

\citep{sun2019bert4rec} builds on the result of SASRec, using the same ID-based approach. The authors propose the BERT4Rec model, which uses a bi-directional attention mechanism rather than a causal one. Hence, they also use a masked language modeling task. The authors show that using this alternative model consistently improves performance across many baselines.

More recent works have looked to alleviate the cold start problem by using language models (LMs) to embed items. For example, \citep{ding2021zero} proposes a model called ZESRec, which uses a pre-trained BERT \citep{devlin2018bert} model to generate item embeddings. The authors then use a Bayesian approach to model the user sequences of item interactions. The trained model can outperform existing methods and generalize across domains. Another related paper proposes the UniSRec \citep{hou2022towards} model, which learns item embeddings by applying a parameterized whitening procedure to BERT embeddings. Finally, \cite{harte2023leveraging} extends the BERT4Rec model to use LLM (i.e., large models) based item embeddings as input. This straightforward change results in 15-20\% improvement on benchmark tasks.

The models discussed thus far use LM embeddings to improve the ability of product recommendation systems to generalize to unseen items and domains. However, these LMs are pre-trained on general natural language rather than the specific data used in each task. \citep{li2023text} proposes formulating the SR problem as a text problem. Their approach, denoted the \textit{Recformer}, constructs item sentences by flattening the key-value pairs into a string and treating everything as text. This model is then trained using the standard masked language modeling objective, along with an additional contrastive task designed to enhance item representations. The Recformer outperforms all other approaches discussed in this section thus far.

Finally, \citep{pancha2022pinnerformer} and \citep{rangadurai2022nxtpost} propose PinnerFormer and NxtPost, respectively, for delivering content recommendations at Pinterest and Meta. In both cases, the recommended items are complex and can comprise images, text, and tabular features. Both papers utilize an event encoding model to embed pins or posts into a latent space, followed by the training of a causal transformer model to predict future behavior. Both of these papers are good examples of how these models can be extended to multimodal domains.

\paragraph{Foundation Models for Tabular Data}
Recent work has explored tabular foundation models, broadly falling into two paradigms \citep{fang2024large}. Serialization approaches convert tabular rows into natural language strings and leverage pretrained language models \citep{hegselmann2023tabllm}. Model-based approaches like FT-Transformer \citep{gorishniy2021revisiting} apply attention mechanisms directly to feature embeddings. However, comprehensive benchmarks show that no single approach consistently outperforms across the broad range of tabular tasks, with gradient boosted trees remaining highly competitive \citep{mcelfresh2023neural}.

Our work addresses a distinct problem: learning representations from time-ordered transaction sequences, not static tabular rows. Within the context of ordered transaction sequences, serialization approaches are unsuitable because text-encoding tabular features is token-intensive, leaving insufficient context for long transaction histories; we compare against text-only serialization in Section~\ref{sec:ablations}. On the other hand, model-based tabular approaches are designed for single-row prediction and do not model dependencies across time-ordered sequences. Rather than replacing tabular models, our Joint Fusion architecture combines learned representations of transaction history with static task-specific tabular features, leveraging the strengths of both.

\paragraph{Sequential Modeling for Financial Data}
Several lines of work apply sequential modeling to financial transaction data. RNN-based approaches include CoLES~\citep{babaev2022coles}, which trains an RNN encoder with a contrastive objective that pulls subsequences from the same user together and pushes apart subsequences from different users, and NPPR~\citep{skalski2023towards}, which trains an RNN autoregressively with a next-item-prediction objective and additionally optimizes the current embedding to predict past behavior. Both report consistent gains over hand-crafted features on banking benchmarks; however, both rely on RNN encoders that struggle with long-range dependencies, and both use only categorical and numerical event attributes, discarding free-text descriptions.

Transformer-based approaches have followed. TabBERT/TabGPT \citep{padhi2021tabular} introduce hierarchical BERT- and GPT-style architectures with field-level transformers over tabular time series. HT-Transformer \citep{karpukhin2025httransformer} addresses transformer under-performance on event-sequence classification by introducing \textit{history tokens} that accumulate context during next-token-prediction pretraining. TransactionGPT~\citep{dou2025transactiongpt} presents a 3D-Transformer foundation model trained on billion-scale payment transactions, supporting downstream classification and generation tasks. A parallel line of work leverages pretrained LLMs directly. LLM4ES~\citep{shestov2025llm4es} textualizes event sequences and fine-tunes an LLM via next-token prediction; LATTE~\citep{fadeev2025latte} aligns raw event embeddings with frozen LLM embeddings via contrastive learning, partly to mitigate the high inference cost of text-based serialization in production.

PRAGMA~\citep{ostroukhov2026pragma} learns representations from heterogeneous banking histories spanning transactions, app activity, trading, and communications. Its input representation separates field identity, value, and temporal information, combining discretized numerical values, categorical tokens, and subword-tokenized text. A hierarchical bidirectional Transformer first encodes each event independently, then contextualizes the event representations alongside the output of a dedicated profile-state encoder. The model is pre-trained through masked reconstruction at the token, event, and field-type levels, jointly learning representations of event histories and profile information. It reports offline evaluations across six banking tasks using linear probes on frozen embeddings or task-specific LoRA fine-tuning of the shared pre-trained backbone.

Our work differs from these in three respects: it operates at production scale (100M+ users, $\sim$500B tokens) with a deployed system; it proposes Joint Fusion to combine learned sequence representations with task-specific tabular features end-to-end rather than producing standalone embeddings; and its compact tokenization is purpose-built to fit hundreds of transactions inside a 2048-token context window, where text-only and hierarchical-token approaches become prohibitively expensive at scale.
\section{A Transformer-Based Model for Transactions}
\label{sec:model_formulation}
Our goal is to ingest a member’s time-ordered transactions and represent their financial behavior as an embedding. Each transaction is represented by text along with numerical and categorical attributes. As is common in other domains like natural language, images, and audio, we show that it is possible to efficiently summarize member behavior by learning to predict their future transactions. 

In this section, we introduce our model formulation based on the transformer \citep{vaswani2017attention}. We choose a causal transformer architecture (GPT-like), as opposed to a BERT-like model or an RNN, for several reasons. First, the state-of-the-art industry approaches to sequential recommendation use a transformer backbone \citep{pancha2022pinnerformer, rangadurai2022nxtpost}. Secondly, transformers offer computational advantages over RNNs \citep{elman1990finding, hochreiter1997long, cho2014learning} during training and inference time, since we are not performing autoregressive generation. We use a causal (rather than masked / BERT-style) objective because next-event prediction aligns naturally with the forward-looking nature of credit risk and matches the dominant approach in production-scale sequential recommendation~\citep{pancha2022pinnerformer, rangadurai2022nxtpost}. Finally, transformers support long range dependencies between inputs. This is especially useful for transaction data because spending money has seasonal variation (e.g. people might spend more money in the holiday season). 

In what follows, we outline the structure of this section: Firstly, in section \ref{sec:model:transaction_interface}, we introduce a modified version of the text-is-all-you-need approach of \citep{li2023text} as our interface between transactions and transformers. Following this, in section \ref{sec:model:finetuning} we discuss a supervised finetuning setup to allow tuning embeddings for specific tasks. Finally, in section \ref{sec:model:joint_fusion} we present an extension of the model, denoted joint fusion, that facilitates end-to-end finetuning with additional tabular features. This is especially important since there are often critical features that are tabular by nature. We denote our joint fusion model \textit{nuFormer}.

\subsection{Transaction Transformer Formulation}
\label{sec:model:transaction_interface}
In this section, we begin by introducing our approach for converting a user, which consists of thousands of transactions, into something that transformer \citep{vaswani2017attention} based sequence-to-sequence models can process (i.e., a sequence of embeddings). Formally, we define a \textit{transaction} as a collection of key-value pairs $t = \{(k_1, v_1), \cdots, (k_m, v_m)\}$. Then, a member consists of a sequence of transactions:
\begin{equation*}
    u_i = \{ t^{(u_i)}_1, \cdots,  t^{(u_i)}_{N_{u_i}}\}.
\end{equation*}

For the purposes of this paper, we will assume a transaction consists of three attributes: the amount represented as a floating point number, the date represented as a timestamp, and finally, a description represented as a string. Hence, in this case each user transaction is defined as:
\begin{equation*}
    t = \{ (\textrm{amt}, v_{\textrm{amt}}), (\textrm{date}, v_{\textrm{date}}), (\textrm{desc}, v_{\textrm{desc}}) \}
\end{equation*}
While this setup is simplistic, we can build representations for the many unique attributes of transactions, such as merchant ID, location, status, merchant category, number of installments, etc.

As discussed, we must first construct an interface between transactions and transformers. One option is to assign IDs to transactions as done in sequential recommendation literature, like SASRec \cite{kang2018self}. However, this faces challenges such as the variability of transaction descriptions over time, leading to a large ID space, and the cold start problem for unseen transactions. Alternatively, we could adopt a text-is-all-you-need \cite{li2023text} approach, converting transaction attributes into a JSON string treated as natural language, which helps in generalizing to unseen transactions. The downside of this method is the large number of tokens it generates per transaction, raising concerns about the quadratic scaling of attention operation costs with context length.

\begin{figure}
    \centering
    \includegraphics[width=\columnwidth]{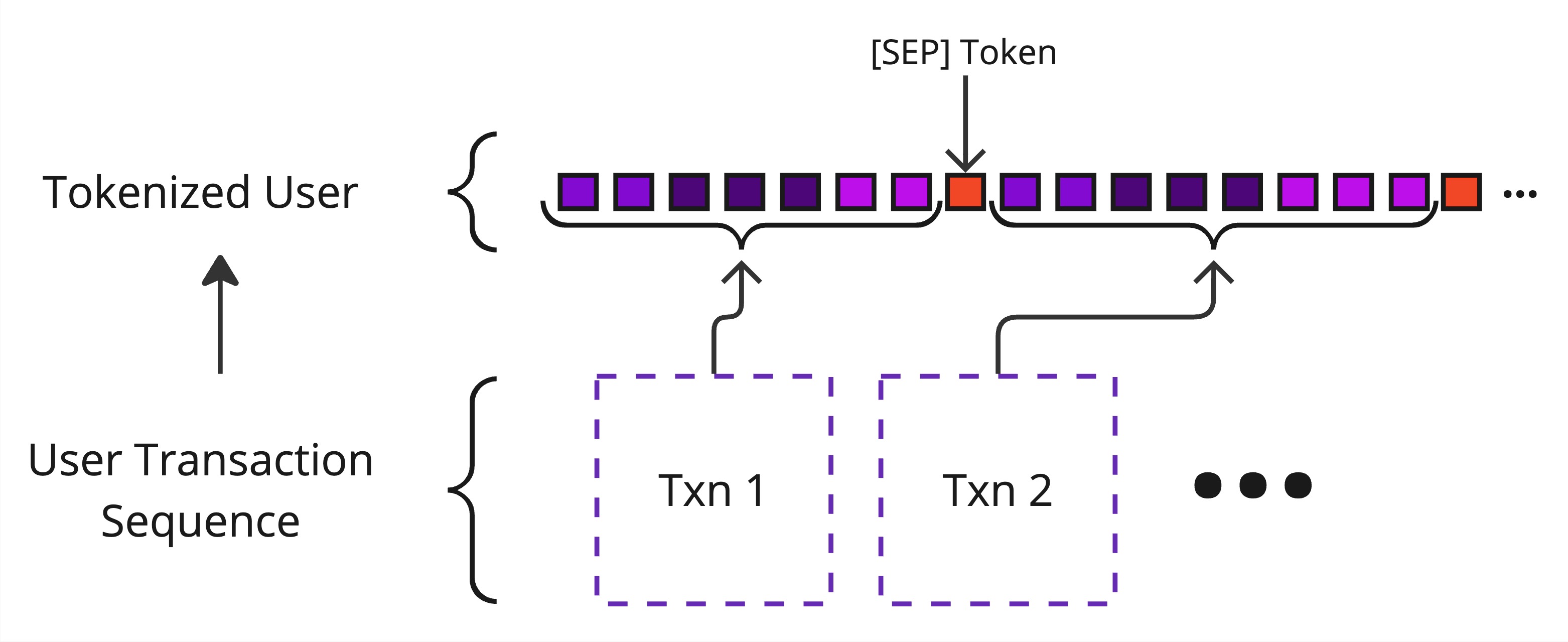}
    \caption{The process of constructing a tokenized member.}
    \label{fig:model:user_tokenization_process}
    \vspace{-13pt}
\end{figure}

For this initial modeling approach, we chose a modified version of text-is-all-you-need. Specifically, we represent numerical and categorical features using special tokens (numerical features are first quantized into bins to make them categorical). Formally, we define a tokenizer as having a vocabulary $\mathcal{V}$ size of $\vert \mathcal{V} \vert = V$ tokens. This vocabulary contains a subset of predefined \textit{special tokens} $\mathcal{V}_S$, with $\vert \mathcal{V}_S \vert = V_S$. The remaining tokens, $\mathcal{V_\textrm{text}} = \mathcal{V} - \mathcal{V}_S$, are for general text and constructed using an algorithm like byte-pair-encoding (BPE) \citep{gage1994new, sennrich2015neural}.
\begin{figure}
    \centering
    \includegraphics[width=\columnwidth]{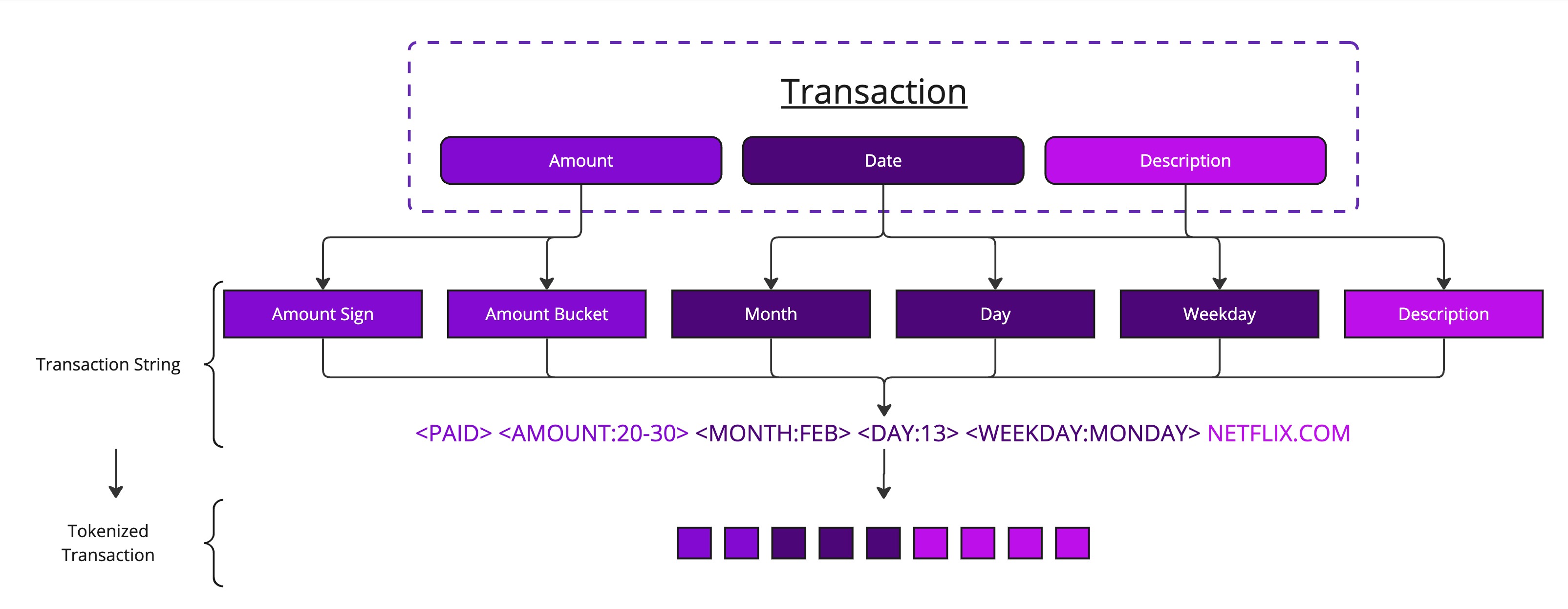}
    \caption{The process of converting a transaction into a stringified and then tokenized form.}
    \label{fig:model:tokenization_process}
    \vspace{-13pt}
\end{figure}
The special token set is crafted from the specific attributes we want to model, specifically the amount, date and description. Specifically, we have the following feature to token mappings:
\begin{itemize}
	\item Amount Sign ($\phi_{\textrm{sign}}: \mathbb{R} \rightarrow \mathcal{V}_{\textrm{sign}}$, where $\vert \mathcal{V}_{\textrm{sign}} \vert = 2$): A separate token to represent whether the transaction amount is an inflow or outflow.
	\item Amount Bucket  ($\phi_{\textrm{amt}}: \mathbb{R} \rightarrow \mathcal{V}_{\textrm{amt}}$, where $\vert \mathcal{V}_{\textrm{amt}} \vert = 21$): The amounts are binned and a separate token is assigned to each bin.
	\item Date Features: are all represented with their own tokens.
        \begin{itemize}
            \item Month  ($\phi_{\textrm{month}}: \mathbb{R} \rightarrow \mathcal{V}_{\textrm{month}}$, where $\vert \mathcal{V}_{\textrm{month}} \vert = 12$): One of the 12 possible months of the year.
            \item Day  ($\phi_{\textrm{day}}: \mathbb{R} \rightarrow \mathcal{V}_{\textrm{day}}$, where $\vert \mathcal{V}_{\textrm{day}} \vert = 31$): One of the possible 31 days.
            \item Weekday  ($\phi_{\textrm{weekday}}: \mathbb{R} \rightarrow \mathcal{V}_{\textrm{weekday}}$, where $\vert \mathcal{V}_{\textrm{weekday}} \vert = 7$): One of the possible 7 week days.
        \end{itemize}
	\item  Text Description ($\phi_{\textrm{BPE}}: \textrm{str} \rightarrow (\tau_1, \cdots), \tau_i \in \mathcal{V_\textrm{text}}$: Tokenized as natural language using a standard tokenizer.
\end{itemize}
where $\mathcal{V}_S = \mathcal{V}_{\textrm{sign}} \cup \mathcal{V}_{\textrm{amt}} \cup \mathcal{V}_{\textrm{month}} \cup \mathcal{V}_{\textrm{day}} \cup \mathcal{V}_{\textrm{weekday}}$. Hence, we can define a tokenized transaction as:
\begin{equation}
\begin{aligned}
    \tau(t) &= (\phi_{\textrm{sign}}(v_{\textrm{amt}}), \phi_{\textrm{amt}}(v_{\textrm{amt}}), \phi_{\textrm{month}}(v_{\textrm{date}}), \phi_{\textrm{day}}(v_{\textrm{date}}), \\
    &\quad \phi_{\textrm{weekday}}(v_{\textrm{date}})) \oplus \phi_{\textrm{BPE}}(v_{\textrm{desc}})
\end{aligned}
\label{equ:tokenized_transaction}
\end{equation}
where $\oplus$ denotes vector concatenation. An example of \eqref{equ:tokenized_transaction} is shown in figure \ref{fig:model:tokenization_process}. Next we can extend this technique to tokenize a member’s account, given by $u_i = \{ t^{(u_i)}_1, \cdots,  t^{(u_i)}_{N}\}$, by concatenating the transaction strings with intermediate separator tokens, $\tau_{\textrm{sep}} \in \mathcal{V_\textrm{text}}$:
\begin{equation}
    \mathbf{x}_i = \left( \bigoplus_{j=1}^{N_i-1} (\tau(t_{i,j}) \oplus (\tau_{\text{sep}})) \right) \oplus \tau(t_{i,N_i}).
\end{equation}
Figure \ref{fig:model:user_tokenization_process} shows this process. 

\begin{figure}
    \centering
    \includegraphics[width=\columnwidth]{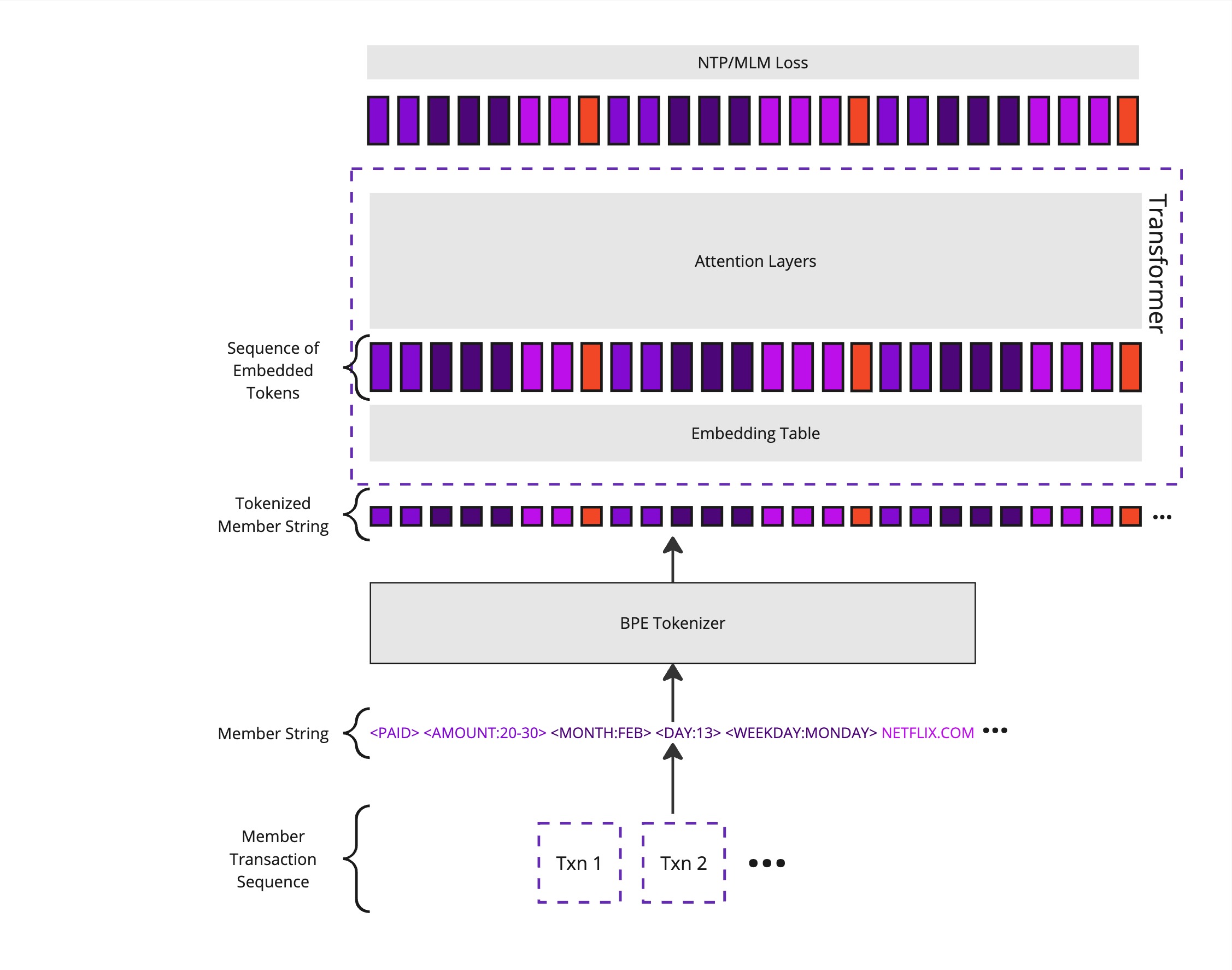}
    \caption{nuFormer pre-training pipeline from raw transactions to next-token prediction.}
    \label{fig:model:user_model}
    \vspace{-13pt}
\end{figure}

Finally, we pre-train causal transformers on our tokenized user representation using the standard next token prediction (NTP) task, which involves classifying the next token to occur out of all possible tokens. This choice is due to the success of standard language modeling tasks in the text-is-all-you-need approach \citep{li2023text}. As with transformers trained on natural language, transaction tokens are embedded using a lookup table. Figure \ref{fig:model:user_model} shows this structure.  

Since the attention operation is invariant to a given token's position in the sequence, it is common to add an encoding of each token's position to the input of the model. \citep{kazemnejad2023impact} showed that causal models learn their own form of position representation. The position information induced by training with a causal mask also generalises to longer sequences during inference \cite{kazemnejad2023impact}. Hence, we choose to use no positional embeddings (NoPE) \citep{kazemnejad2023impact}. Finally, we also use FlashAttention \citep{dao2022flashattention, dao2023flashattention, shah2024flashattention}. Both NoPE and FlashAttention allow us to train on large context lengths within a single A100 80GB GPU. In practice, we often train on clusters of H100 or H200 GPUs using distributed data parallel or fully sharded training.

\subsection{Finetuning Embedding Models}
\label{sec:model:finetuning}
In the previous section, we introduced a formulation of transaction sequences as natural language. This facilitated pre-training user embedding models on transaction data, and learning general embeddings of user behavior. In practice, however, such models are often fine-tuned for specific tasks to achieve state-of-the-art performance. While in this section, we consider finetuning our models for binary classification problems, the same technique can be used for multi-class or even regression problems.

During supervised finetuning, a member consists of a sequence of transactions: $u_i = \{ t^{(u_i)}_1, \cdots,  t^{(u_i)}_{N_{u_i}}\}$, and a label $l_i$. We wish to learn a function $f_\theta(u_i) = l_i$. To achieve this, we take the final token (before any padding) embedding as the user representation, and add an MLP network to the model, which reduces this embedding to a score. We choose the final token embedding because, since the model is causal, it is the only token with the full context of the user. Then, we optimize the MLP and the transformer weights to minimize the cross-entropy error. 

Full fine-tuning of the transformer is memory-intensive, particularly when training jointly with the DCNv2 tabular network. To make training with Distributed Data Parallel (DDP) on our GPU resources feasible, we use LoRA \citep{hu2021lora}, which freezes the pre-trained weights and injects trainable low-rank matrices into each transformer layer. This reduces memory requirements substantially while retaining most of the performance benefit of full fine-tuning (see Section~\ref{sec:ablations}).

\subsection{Modeling Tabular Features with DNNs}
\label{sec:model:joint_fusion}
A motivation for transaction-based user embedding models is to alleviate the need for manual feature engineering. In the previous section, we saw how to refine the transaction embeddings for specific tasks through finetuning. However, in many cases, there are either hand-crafted features from non-transaction sources or features that are tabular by nature that are critical to model quality. Since these handcrafted features are orthogonal to the transaction data used to train the embedding models, we need to incorporate them into the final prediction. This process of combining embeddings with tabular features is denoted \textit{blending} or \textit{fusion}.

Gradient-boosted tree (GBT) models are generally considered state-of-the-art for dealing with tabular data \citep{borisov2022deep}, e.g., XGBoost \citep{chen2016xgboost} or LightGBM \cite{ke2017lightgbm}. There are two common approaches for blending with GBTs. The first is early fusion, which blends pre-trained embeddings with tabular features using GBTs. The second is late fusion, a two-stage process, where we first finetune the transformer on a subset of the data and then train the GBT model to combine the finetuned embeddings with tabular features. 

\begin{figure}
	\includegraphics[width=0.99\columnwidth]{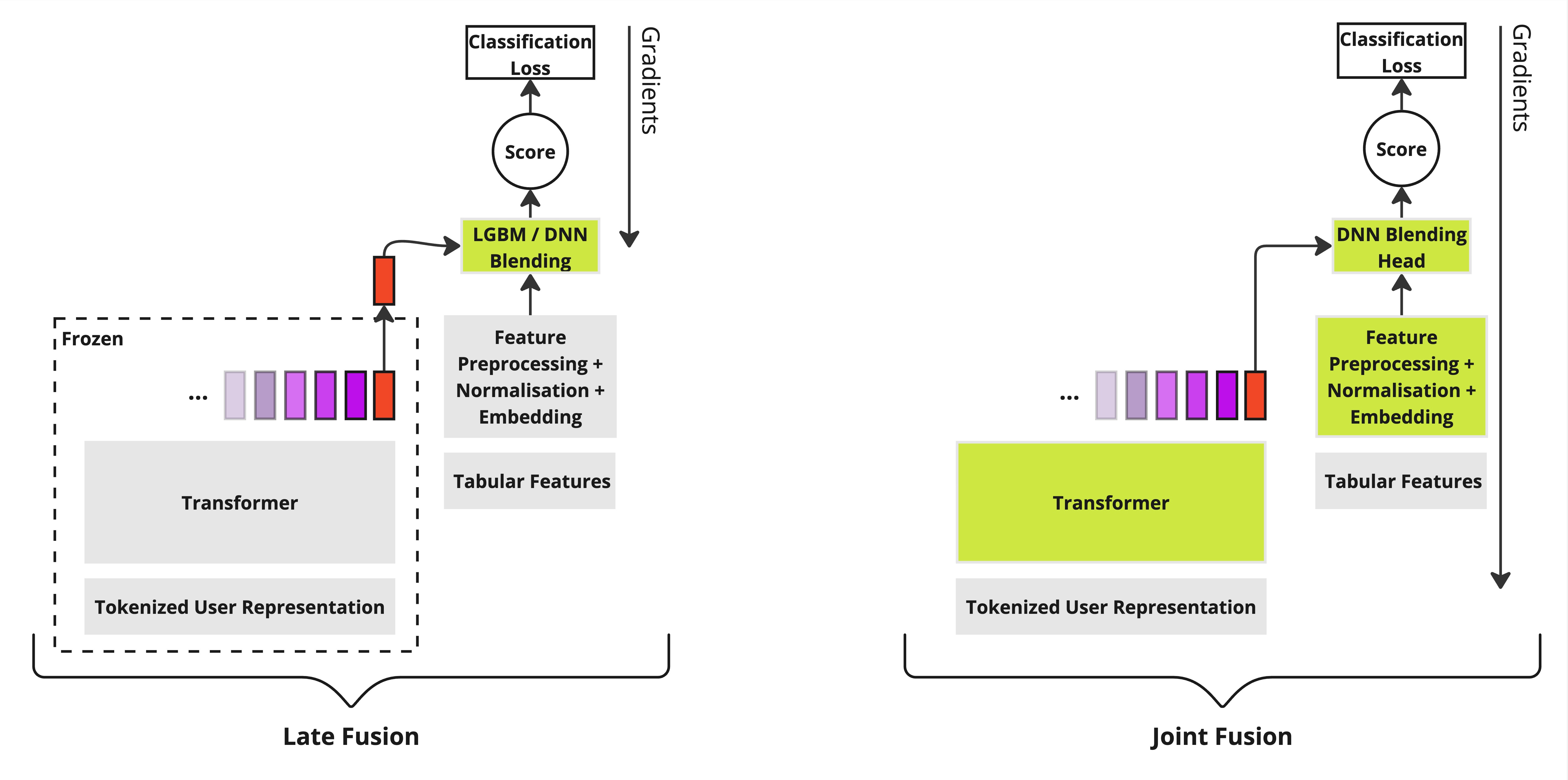}
	\caption{The difference between late fusion and joint fusion (green boxes indicate learnable parameters).}
	\label{fig:model:late_fusion_vs_joint_fusion}
    \vspace{-13pt}
\end{figure}

Late fusion allows for embeddings that are tailored to the task of interest and performs better than early fusion. However, these finetuned embeddings are learned in isolation from the features. We hypothesize that finetuning jointly with blending will allow the transformer to better capture interactions between the tabular features and embedded transaction data. Hence, in contrast to both early/late fusion, we propose a system that can be trained end-to-end, allowing the model to learn an optimal blending of tabular and sequential data, which we denote as joint fusion. Figure \ref{fig:model:late_fusion_vs_joint_fusion} shows the joint and late fusion systems.

GBTs are not compatible with joint fusion because there are no gradients from the GBT to propagate to the transformer. Motivated by these observations, we invested in DNN-based tabular feature networks, for which recent works have started to show they can be competitive with GBTs \citep{zabergja2024tabular}. However, the performance of DNN tabular feature networks can vary drastically between problems. For example, one survey paper \citep{mcelfresh2023neural} evaluated 19 different tabular feature models (NN + GBT) on 176 distinct datasets. The authors found that each of the 19 different models outperformed all others on at least one dataset. On the other hand, there also existed at least one dataset where each model simultaneously performed the worst. Hence, making it challenging to have a one-size-fits-all approach. Therefore, we need to find a configuration that works for our problems.

The first step in our approach was to achieve parity between DNNs and GBT models on only the tabular features. We selected the DCNv2 architecture \citep{wang2021dcn} as it has shown success on related problems at a large scale (e.g., used by Google). However, initial results showed much worse performance for the DNN-based DCNv2 models than for the GBTs.

The recent paper \citep{gorishniy2022embeddings} found that by embedding numerical attributes, they achieved significant gains when modeling numerical tabular features in DNNs. These numerical embeddings are constructed using periodic activations at different (learned) frequencies. We combined this with trainable embedding tables to also facilitate categorical feature embeddings. Combined with L2 regularization on the cross layers, this DCNv2 + PLR configuration matches and slightly exceeds the LightGBM baseline (+0.08 \%) on tabular features alone, giving us a DNN tabular branch suitable for Joint Fusion. Appendix~\ref{sec:results:ablation_of_dnn} reports the full ablation across MLP, DCNv2, and embedding variants.

Despite achieving parity with only tabular features, the last challenge to overcome was incorporating embeddings into these models while maintaining or beating the GBT model's performance with DNNs. Three key factors were critical in achieving this. First, we implement modifications to the DCNv2 architecture for our use case, utilizing its cross-layer capabilities to process only the embedded tabular features and project the result into a low-dimensional embedding. This feature embedding is concatenated with the transaction embedding (from the transformer), and a multi-layer perceptron is used to make the final prediction. Secondly, adding regularization in the form of weight decay and/or dropout to the customized DCNv2 layers reduced overfitting. Finally, adding normalization to the transaction embeddings improved the consistency of the customized DCNv2, allowing the DNN model to outperform the GBTs reliably. Despite the challenges in using DNNs with tabular data, we achieved a model that works well for our current tasks of interest by combining DCNv2, numerical, and categorical feature embeddings, along with regularization. We denote this joint fusion model as \textit{nuFormer}.

In summary, Joint Fusion required overcoming several key challenges. Specifically, replacing GBTs with neural networks that can propagate gradients, achieving DNN-GBT parity through the aforementioned careful architectural choices, and managing the memory and compute costs of jointly optimizing a 330M-parameter transformer with DCNv2. We address the latter two through LoRA \citep{hu2021lora} for memory-efficient fine-tuning and a two-stage hyperparameter search that reduces cost by approximately 100$\times$ (see Appendix~\ref{sec:appendix}).
\section{Experimental Results}
\label{sec:experimental_results}
In this section, we empirically evaluate our transformer-based embedding model for transaction data on a practical credit default prediction task. First, in section \ref{sec:results:problem_and_baseline} we introduce the credit default problem in more detail as well as the LightGBM baseline. Then, in section \ref{sec:results:ablation_of_foundation_models} we explore how pre-training and joint fusion scale as a function of several model properties (model size, context length, data volume). Next, in section \ref{sec:results:end_to_end_modeling}, we apply our transformer-based embedding models to a practical credit default modeling task using backtest data. Importantly, we show that we can achieve a 1.25\% improvement in test set AUC by using our transformer models; this lift in performance is $3$x a typical model launch that leads to a significant business outcome. This section includes a discussion of our production deployment of these models to over 100M+ users. Following this, in section \ref{sec:extended_offline_results} we show the ability of this approach to work across tasks and other regions. Finally, in section \ref{sec:ablations}, we present an ablation of several aspects of the nuFormer model, including a comparison against a text-only model.

\subsection{Credit Default Prediction Problem and Baseline Setup}
\label{sec:results:problem_and_baseline}
In this section, we introduce the credit default prediction problem that we use in following sections to demonstrate the success of our representation learning models. The data consists of 203M training rows and 2M testing rows, where each row corresponds to a particular label and timestamp combination for a given user. Specifically, the same user might occur multiple times in the dataset, but at different times, with potentially different labels and/or transaction sets. The label is binary and represents whether user being 65 days late on a payment 6 months from the label date (1 yes, 0 no). Importantly, this label is time-delayed, which means we are attempting to predict the user behavior in six months from the score date. The time periods covered by the train and test sets are disjoint.

Each row can contain potentially many transactions, though in some cases, members might have no transaction history. For these experiments, the transactions of a member can be generated from three independent financial products. The specific products don't matter as much for the experiment so we denote them as sources A, B and C. In Appendix~\ref{sec:transformer_training:data_source_comb}, we ablate combinations of these sources via early fusion: source A provides the strongest single signal, source B adds orthogonal information, and source C can dilute performance by displacing higher-signal transactions within the context window. Our main experiments below use all three sources.

In total there are 291 tabular (numerical or categorical) features. Some are derived from transaction sources (e.g., average spend over a given period), but non-transaction sources such as bureau scores are also included. Unless otherwise noted, all improvements are reported as relative improvement (\%) in AUC over a LightGBM model trained on tabular features only, which is our production baseline.

Our joint fusion model learns features that make some hand-crafted transaction features redundant. Although not all transaction sources used to construct these features are currently provided to the model as raw transactions, in practice we can remove 90\% of the redundant hand-crafted transaction features with no loss in performance.

\subsection{Exploring Pre-Training and Finetuning of Transformer-Based User Embedding Models}
\label{sec:results:ablation_of_foundation_models}

This section presents an analysis of how the embeddings of our finetuned LMs improve in quality as we scale and various other ablations. To begin, we define the baseline models, of which there are two, with 24M and 330M parameters, respectively. Both have a context length of 2048 and use all the transaction sources. These models are pre-trained and finetuned on 20M rows. The baseline transformers take advantage of description, amount, and date, as described in figure \ref{fig:model:tokenization_process}. All subsequent experiments use joint fusion, and results are reported as the relative improvement in AUC over a GBT baseline trained only on tabular features using all 203M rows. 
In section \ref{sec:transformer_training:model_size} we show that using larger models allows us to learn better features from the raw transaction data. Then, in section \ref{sec:transformer_training:context_length}, we examine how varying the context length affects the performance of joint fusion. Finally, in section \ref{sec:transformer_training:data_volume}, we explore how the volume of finetuning data affects the downstream performance.

\subsubsection{Impact of Model Size}
\label{sec:transformer_training:model_size}

We compare the AUC performance of finetuned models of two different sizes: one model consists of approximately 24M parameters while the other has approximately 330M parameters. We use a causal GPT-like decoder-only model, with 24 attention layers, each with 16 attention heads. Importantly, the two model configurations differ only in the hidden embedding size where the 330M model has a 1024 length embedding and the 24M model has a smaller 256 length embedding. 

The 24M and 330M models achieve a lift of \textbf{0.37\%} and \textbf{0.62\%} over the baseline model configuration, respectively. We see an improvement from the increased capacity. The findings in this section suggest that there is a rich set of information in our data that larger models can exploit. In future work, we plan to continue experimenting with scaling up to larger models, and derive scaling laws for these transaction based user embedding models.

\subsubsection{Effect of Context Length on Joint Fusion}
\label{sec:transformer_training:context_length}

\begin{figure}
    \centering
    \includegraphics[width=1.0\linewidth]{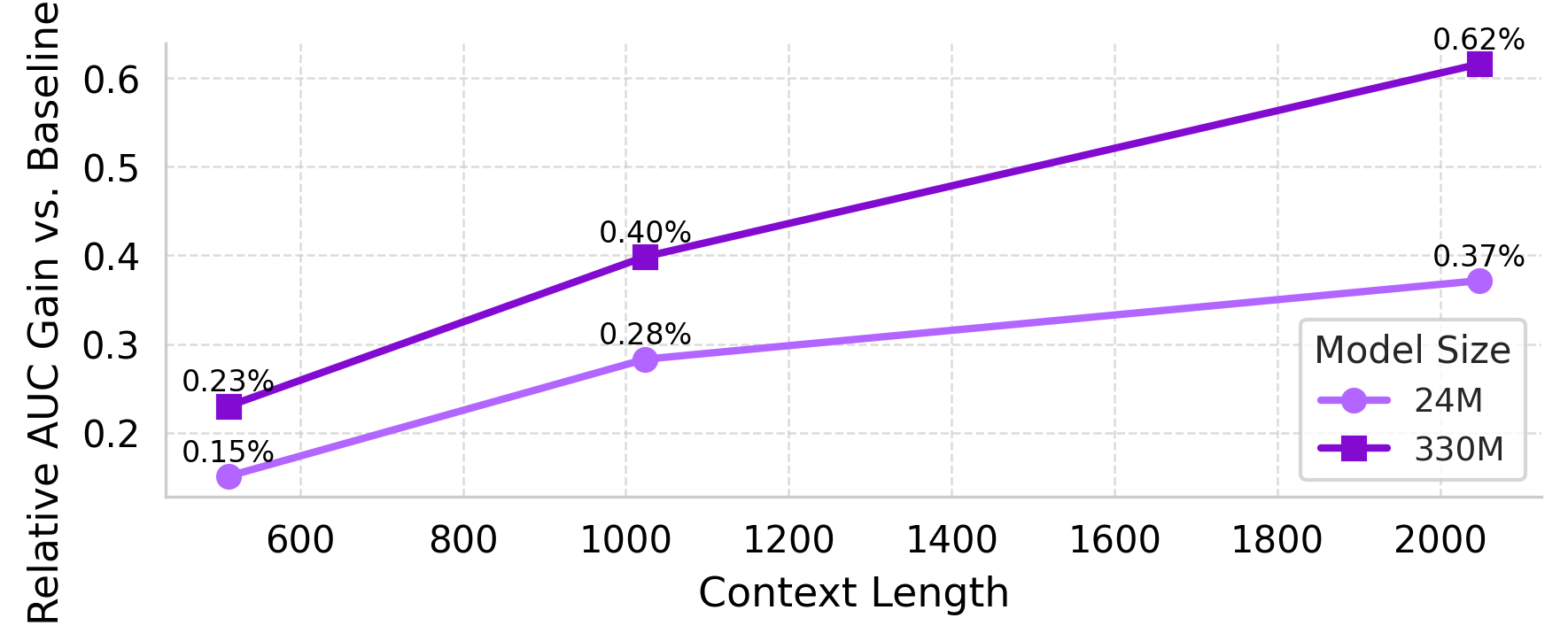}
    \caption{\textit{nuFormer} relative test AUC gain for different context lengths (512, 1024, 2048) with 24M and 330M models.}
    \label{fig:jf_test_auc_vs_jf_context_length}
    \vspace{-13pt}
\end{figure}

In this section, we explore how varying the context length affects the performance of joint fusion. More specifically, we compare the baseline configuration using context lengths of 512, 1024 and 2048 tokens for both the 24M and 330M variants. Figure \ref{fig:jf_test_auc_vs_jf_context_length} shows these results, where we see that larger context lengths lead to improved performance. Moreover, we see that the larger model is better able to exploit the information in the longer contexts. 

\subsubsection{Effect of Training Data Volume for Joint Fusion}
\label{sec:transformer_training:data_volume}

\begin{figure}
    \centering
    \includegraphics[width=1.0\linewidth]{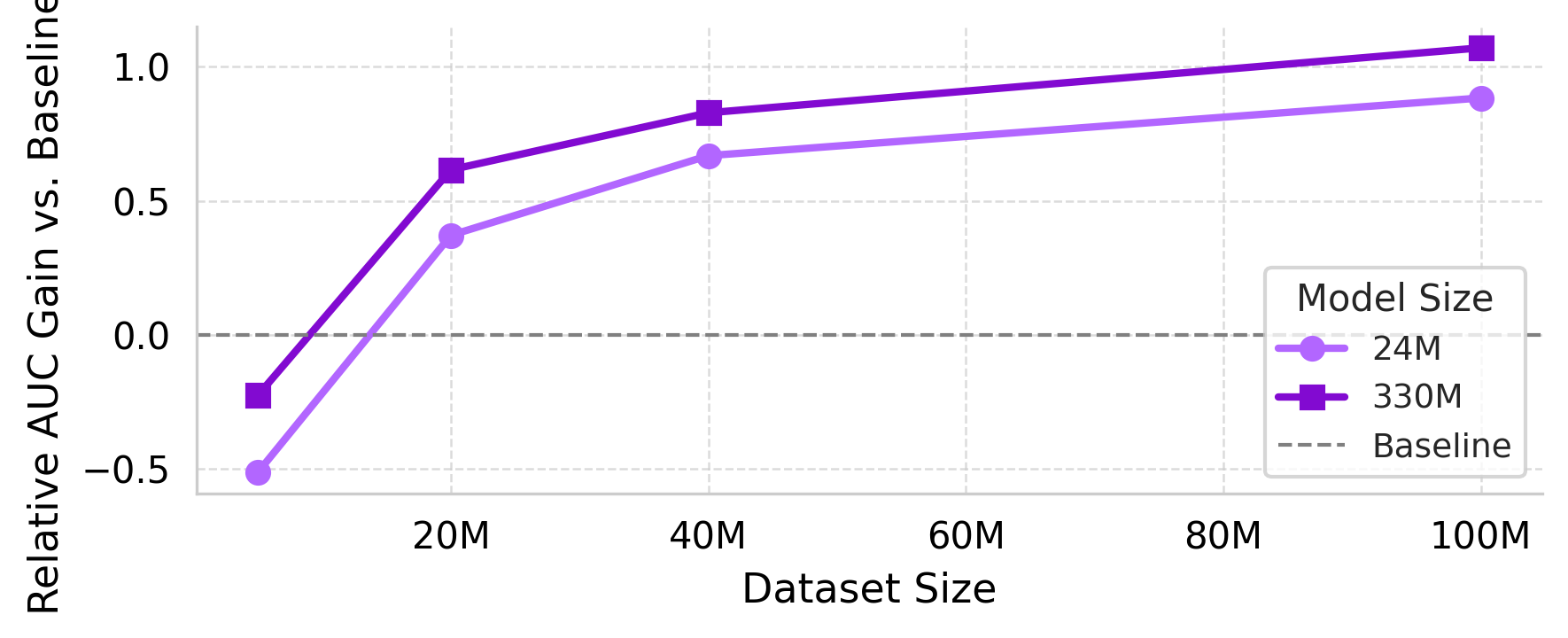}
    \caption{\textit{nuFormer} relative test AUC gain for different amounts of training data (5M, 20M, 40M, 100M) with 24M and 330M models.}
    \label{fig:jf_test_auc_vs_jf_data_volume}
    \vspace{-13pt}
\end{figure}

In this section, we demonstrate that the performance of our joint fusion scales as a function of the data volume. Specifically, we take two models, one 24M and one 330M, both pre-trained on 20M rows. We then finetune each of these two models on 5M, 20M, 40M, 100M rows, and plot the results, which is shown in figure \ref{fig:jf_test_auc_vs_jf_data_volume}. Importantly, we see a clear advantage to using more data with joint fusion, and that bigger models obtain a larger advantage from more data.

\subsection{Production Deployment for Credit Default Prediction}
\label{sec:results:end_to_end_modeling}
This section presents a practical application of our transaction data model to a credit default prediction problem. This problem involves predicting whether a member default on a loan if given a credit limit increase. The baseline for this task attempts to predict this behavior from hand-crafted tabular features using LightGBM. Importantly, some of these tabular features are derived from transactions. Hence, our transaction embeddings do not add any new sources of data to the model. Rather, we are allowing the model to learn its own transaction features from the raw data.

We compare several different strategies to clearly demonstrate the advantage of our final Joint Fusion approach, which achieves an overall relative improvement of $+1.25\%$ in test AUC over the baseline model trained only on the tabular features. For reference, this is $3$x the improvement typically observed when successful models realize material business impact. Furthermore, historical model improvements were obtained from adding new data sources or signals to the model. In our case, this improvement is entirely from advancements in modeling.

The data consists of one record per user interaction. Hence, the total amount of data is much greater than just one row per member. In the following experiments, we want to analyze the gain in test AUC from adding the transformer model user embeddings. We blend this model with the tabular features in the following ways:
\begin{enumerate}
	\item Baseline: The existing model uses the handcrafted tabular features and is a LightGBM trained on the entire dataset. This model has no user embeddings.
	\item Late Fusion: We finetune the model on a 20\% subset of the data rows without any features, and then we use the remaining rows to train the downstream models to predict the label from the embeddings and features. 
	\item \textit{nuFormer}: Uses the joint fusion procedure on 203M rows.
\end{enumerate}
Table \ref{tab:product_recomendation_results} shows the improvements over the baseline in the test AUC from adding transaction based user embeddings.

\begin{table}[t]
\begin{tabular}{l | l}
	\toprule
	Model & Relative AUC Improvement \\
	\midrule
	Late Fusion (LightGBM) & 0.97\% \\
	Late Fusion (DCNv2) & 0.97\% \\
	\textit{nuFormer} & 1.25\% \\
	\bottomrule
\end{tabular}
\caption{Relative AUC Improvement (\%) over the baseline (LGBM) for different blending approaches.}
\label{tab:product_recomendation_results}
\vspace{-15pt}
\end{table}

\begin{figure}
    \centering
    \includegraphics[width=0.9\linewidth]{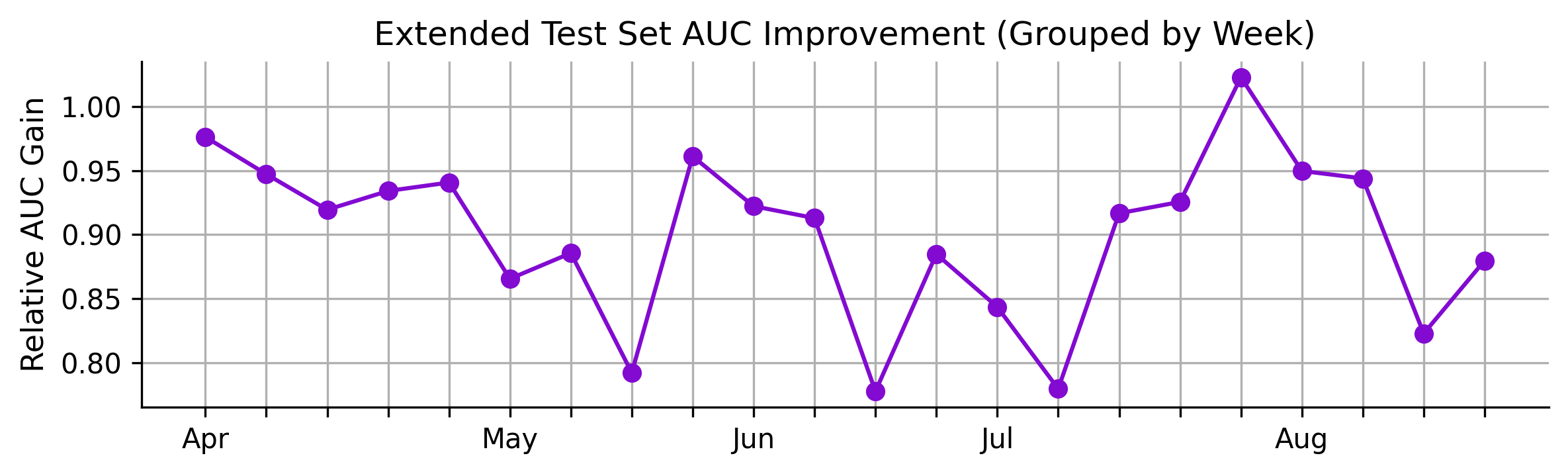}
    \caption{Extended out-of-time test-set stability analysis, comparing \textit{nuFormer} against the baseline.}
    \label{fig:extended_test_set_stability_analysis}
    \vspace{-13pt}
\end{figure}

A major concern was whether the joint fusion based solution would overfit to transaction attributes (e.g., descriptions) from specific time periods. This is especially important because labels take six months to mature, so we cannot train these models on the most recent transactions. During the back-testing, we constructed an extended test set, covering a 6 month period after the data used during joint fusion. Figure \ref{fig:extended_test_set_stability_analysis} shows the difference in AUC for this extended test set. Importantly, we see a consistent gain over the baseline as we get further from the train period.

The nuFormer model (Joint Fusion with DCNv2) was deployed to production with the primary goal of reducing long-term user defaults. The system's success is measured by the reduction in users being late on payments. The long lead-up to collecting the label required for the business outcome prevented us from A/B testing every variant. Hence, nuFormer is currently deployed to 100\% of customers in one of our markets, serving credit decisions for over 100 million users. The remainder of this section describes the infrastructure, data integrity safeguards, and monitoring systems required for production operation.

\paragraph{Inference Infrastructure.} Inference runs as a daily batch job across the full customer base. Processing 100M+ users typically requires approximately 5 hours on 60 A100 GPUs (or 150-300 L4 nodes). With a batch size of 22 on A100s, the model consumes roughly 90\% of available GPU memory, reflecting the cost of long-context transformer inference at scale.

\paragraph{Data Integrity.}
Financial transactions are mutable: amounts, descriptions, and status fields can change as transactions move from pending to approved to reversed. If models observe updated values that were unavailable at label generation time, data leakage occurs. We implement two safeguards. Firstly, for training and inference, transactions are presented as they existed at the score date, ensuring the model only accesses information available at decision time. Secondly, automated systems compare current records against historical snapshots to detect unintended field changes. This uncovered an issue where Buy-Now-Pay-Later prepayments altered transaction timestamps, leaking future information into the present.

\paragraph{Drift Monitoring.}
We continuously compare recent data against training-period distributions to detect shifts that could degrade model performance. Firstly, we examine field-level drift by monitoring every input field (e.g., description text, amounts, timestamps) for distributional changes. Secondly, we examine behavioral drift by tracking shifts in usage patterns across aggregate metrics (daily transaction counts, transactions per user, total spend).

\paragraph{Online Performance.}
The deployed model 
drove a 32\% increase in our long-term surrogate metric, a contribution equivalent in magnitude to an entire year's typical cumulative gains and approximately $3\small\times$ the impact of a standard modeling upgrade. Crucially, these improvements came purely from enhanced representation learning on existing transaction data, without adding new data sources or manual feature engineering. Figure~\ref{fig:extended_test_set_stability_analysis} shows that performance gains remain stable across a 6-month out-of-time evaluation period, indicating the model does not overfit to transaction descriptions or temporal artifacts from the training period.

\subsection{Generalizing the nuFormer Paradigm Across Tasks and Markets}
\label{sec:extended_offline_results}
To evaluate whether nuFormer's methodology generalizes beyond the primary credit task, we applied the same approach to four additional problems across markets. Each model follows the same architecture and training procedure: pre-training a transaction transformer on local transaction data, followed by Joint Fusion fine-tuning for the target task. Table \ref{tab:extended_offline_results} reports the offline results.

\begin{table}[ht]
\centering
\small
\begin{tabular}{@{}llcr@{}}
\midrule
Task & Market & Metric & Gain \\
\midrule
Credit default (primary) & Market 1 & AUC & +1.25\% relative \\
Credit default & Market 2 & AUC & +1.81\% relative \\
Lending & Market 1 & AUC & +2.23\% relative \\
Income prediction & Market 1 & F1 & +7.54\% relative \\
Spend prediction & Market 1 & R\textsuperscript{2} & +6.85\% relative \\
\midrule
\end{tabular}
\caption{nuFormer improvements across tasks and markets vs.\ baseline models, reported as ordinary relative lift.}
\label{tab:extended_offline_results}
\vspace{-20pt}
\end{table}

These results demonstrate that the nuFormer methodology transfers across diverse prediction tasks, markets, and problem types (classification and regression) without architectural modifications. The Market 2 credit model, pre-trained entirely on transaction data from that market, achieves gains comparable to the primary deployment, suggesting the approach is not specific to a single market's transaction patterns. Each model presented in Table \ref{tab:extended_offline_results} is currently deployed in production and serving users.

\subsection{Component Ablation Study}
\label{sec:ablations}

The preceding experiments demonstrate that Joint Fusion improves credit prediction across markets and tasks. We now analyze which components drive these gains. Due to computational constraints, the following ablations use a 10M row benchmark subset with 1024 context length. Table~\ref{tab:ablations} reports results across four dimensions: modality (what data the model sees), training strategy (how components are optimized), and tokenization (how transactions are represented). For the tokenization comparison, we also vary context length to control for the number of transactions observed, isolating the effect of representation choice from context efficiency.


\begin{table}[ht]
\centering
\small
\setlength{\tabcolsep}{4pt}
\begin{tabular}{@{}lccc@{}}
\midrule
Model & Ctx & Txns & Relative Gain \\
\midrule
\multicolumn{4}{@{}l}{\textit{Modality}} \\
Tabular-only & — & — & -0.96\% \\
Transformer-only & 1024 & 81 & -2.83\% \\
\midrule
\multicolumn{4}{@{}l}{\textit{Training Strategy}} \\
Early Fusion & 1024 & 81 & -0.51\% \\
Joint Fusion & 1024 & 81 & \textbf{—} \\
\midrule
\multicolumn{4}{@{}l}{\textit{Tokenization (matched txn count)}} \\
Joint Fusion & 352 & 28 & -0.36\% \\
Text-only (scratch) & 1024 & 31 & -0.37\% \\
Joint Fusion & 224 & 18 & -0.45\% \\
Text-only (scratch) & 662 & 20 & -0.54\% \\
Text-only (GPT-2 PT) & 1024 & 18 & -0.91\% \\
Text-only (Qwen 0.6B) & 1024 & 24 & -0.41\% \\
Text-only (Qwen 4B) & 1024 & 24 & -0.43\% \\
\midrule
\end{tabular}
\caption{Ablation on 10M rows. Relative gain is reported as ordinary relative change in AUC versus the Joint Fusion baseline. Joint Fusion uses special tokens; Text-only (scratch) uses domain BPE; GPT-2 PT/Qwen use pre-trained tokenizers.}
\label{tab:ablations}
\vspace{-20pt}
\end{table}

\paragraph{Both modalities contribute.} Tabular only (-0.96\%) outperforms Transformer only (-2.83\%), indicating that hand-crafted features remain valuable for credit prediction. However, combining both modalities via Early Fusion (-0.51\%) exceeds either alone, demonstrating that transaction embeddings and tabular features capture complementary signals. The transformer learns patterns from raw sequences that are not fully captured by aggregated statistics.

\paragraph{Joint training outperforms early fusion.} Joint Fusion outperforms Early Fusion by 0.51\% despite using identical inputs. End-to-end training enables the model to learn interactions between transaction patterns and tabular features that cannot be captured when representations are simply concatenated. This validates our architectural choice of jointly optimizing the transformer and DCNv2.

\paragraph{Naive Usage of World Models is not Advantageous} We evaluated open-source pre-trained models within Joint Fusion: a GPT-2 Medium model pre-trained on Wikipedia, architecturally identical to our from-scratch models, and the Qwen3 0.6B and 4B models \citep{yang2025qwen3}. Text-only GPT-2 PT achieves -0.91\%, underperforming both Text-only scratch (-0.54\%) and Joint Fusion scratch (-0.45\%) despite access to pre-trained language understanding. The more modern Qwen3 models perform better, 0.6B and 4B achieve -0.41\% and -0.43\%, respectively, but still fall below our from-scratch Joint Fusion baseline while requiring 3.7$\times$ and 9$\times$ the training time due to larger model and vocabulary sizes.

These results do not preclude the value of world knowledge for transaction modeling, but suggest that naive application of pre-trained language models is insufficient. Promising directions for future work include domain-adaptive pre-training on transaction corpora, optimizing the text-only representation of transaction data, developing tokenizers optimized for financial text, optimizing prompts to improve the presentation of transaction data to the model, and hybrid approaches such as semantic IDs \citep{rajput2024recommender}.

\paragraph{Controlled comparison by transaction count.}
To isolate the effect of tokenization from context efficiency, we compare models at matched transaction counts by varying context length. At approximately 28 to 31 transactions, Joint Fusion (352 context, -0.36\%) performs comparably to Text-only scratch (1024 context, -0.37\%). At approximately 18 to 20 transactions, Joint Fusion (224 context, -0.45\%) slightly outperforms Text-only scratch (662 context, -0.54\%). These results suggest that, under equal behavioral context, the two representations achieve similar performance. However, the advantage of compact tokenization is primarily efficiency, enabling 3$\times$ more transactions within the same context window.

\section{Conclusion}
\label{sec:conclusion}
In this paper, we introduced a novel approach to leveraging transformer based embedding models for financial data, transforming raw transactions into actionable insights. While these models build on standard data sources used throughout the industry, they facilitate automatically learning informative features that may not be obvious to data scientists. In an empirical evaluation, we saw how using joint fusion to tune our transaction-based transformer models for a credit default prediction task could generate substantial lifts both offline and online, deployed to over 100M+ customers. These models can be leveraged for tasks across markets, improving our ability to understand our customers so we can help them meet their financial needs at the right time. Moreover, we saw a clear advantage to scaling the model size, context length, and training data. In future work, we plan to develop scaling laws for these models and further explore exploiting world knowledge.

\bibliographystyle{ACM-Reference-Format}
\bibliography{paper}

\appendix
\section{Data Source Combinations}
\label{sec:transformer_training:data_source_comb}
Our users can have multiple types of accounts, cards, and transactions within their event history, which can be broken down into numerous data sources, including credit card, debit card, open finance, wires, transfer activity, and a variety of bill items like credit card payments and fees. Importantly, when using the tokenization procedure described in figure \ref{fig:model:tokenization_process}, each transaction consumes 14 tokens on average. Hence, given that our LMs have a limited capacity and context window, it is important to explore the power and impact of each individual data source to find the blend and balance for optimal performance. 

We show that even though every source of data adds a different view of the consumer's activity, they might not all incrementally add value to the predictive power of the model. To study the effect of each transaction set, we pretrained numerous 24M models with various combinations of $3$ anonymized data sources ($A$, $B$, $C$), extracted those pretrained embeddings, and trained a GBT to make predictions based on the information stored in the pretrained embeddings (i.e, early fusion). Importantly, table \ref{tab:sources} presents an ablation study comparing transaction source combinations. Unlike other tables, the baseline here is early fusion using all sources (ABC), as the goal is to understand the relative importance of each source rather than improvement over production. All other tables report improvements to the production LightGBM baseline trained on tabular features only. Overall, data source $A$ clearly contains powerful information relevant for credit default. Transactions from $B$ also contain useful signal that appears to be orthogonal to $A$, but $C$ may slightly confuse the model or take attention from more pertinent transactions when combined with other sources.

\begin{table}[h!]
\centering
\begin{tabular}{@{}lr@{}}
Source & Relative AUC Gain vs.\ ABC (\%) \\
\midrule
A   & +3.92\% \\
B   & -44.67\% \\
C   & -111.64\% \\
AB  & \textbf{+4.95\%} \\
BC  & -66.59\% \\
AC  & -1.47\% \\
ABC & 0.00\% \\
\end{tabular}
\caption{Relative AUC gain versus the all-sources baseline (ABC) for different source combinations, computed as shortfall reduction.}
\label{tab:sources}
\vspace{-20pt}
\end{table}

Interestingly, in some cases, adding a data source can cause a decrease in performance. For example, $BC$ < $B$ and $ABC$ < $AB$. This is caused by the additional transactions causing increased contention in the already limited context window. In this case,  transactions from $B$ appear to contain the least useful information, and when we include them, it pushes more useful transactions from $A$ and $C$ out of the model's visibility. 

This aforementioned effect can be caused by the difference in information densities and transaction frequencies between sources. In some cases, transaction sources can be sparse but relevant. For example, bill items can contain information about buy-now-pay-later loans. On the other hand, debit card can contain many (frequent) small transactions which are overall less important.

\section{Tabular Feature Modeling Parity with DNNs}
\label{sec:results:ablation_of_dnn}

In this section, we explore the aspects of the DNN model that allow us to match the LightGBM performance when modeling the tabular features (numerical and categorical) highlighting the incremental gains obtained towards the challenger DNN model. Table \ref{tab:dnn_ablation_results} shows the improvement over the baseline (LightGBM) as we improve our DNN tabular feature model. It is important to highlight that the results shown here for each DNN and LightGBM model are obtained via hyperparameter tuning. See Appendix~\ref{sec:appendix} for complete architecture and hyperparameter details.

The first model was a multilayer perceptron architecture (MLP), preceded by a processing step that applies standardization to numerical features and one hot encoding to categorical features. Table \ref{tab:dnn_ablation_results} shows that the MLP approach was unable to match the performance of LightGBM. The second was the Deep Cross Network V2 (DCNv2) architecture designed by Google \citep{wang2021dcn}, which is capable of modeling explicit and implicit interactions of input features to make the final prediction. For this particular network, a different feature processing is employed: numerical features are transformed through a signed \emph{log1p} function and categorical features are mapped to learnable embeddings via lookup table. These transformations were necessary to avoid numerical instability in the cross layers, as they do not work with sparse vectors.

Although DCNv2 achieved higher performance than MLP, it still lagged behind LightGBM baseline. Recent work \citep{gorishniy2022embeddings} has shown that representing numerical attributes as dense embeddings helps to improve the performance of neural networks on tabular-based problems, either for MLP or Transformer based architectures. Following that idea, we retrained both MLP and DCNv2 models using the periodic linear (PLR) embedding approach for numerical features, which maps a real number to a dense embedding of parametrized size (number of frequencies), whose elements are periodic activations (sin and cosine) of the numerical value. Results on Table \ref{tab:dnn_ablation_results} show the effectiveness of this embedding technique: MLP and DCNv2 with numerical periodic embeddings outperform their previous performances, and DCNv2 + periodic embeddings was able to match the performance of LightGBM.

The use of numerical embeddings usually leads to an increase in the number of parameters of DNNs because the input vector becomes wider (depending on the dimension of the embedding). Naturally, the model becomes more prone to overfit and requires regularization to avoid losses in generalization power. Thus, in a following step, we applied L2 regularization to DCNv2 weights, which further increased performance on the test set, furthering the DCNv2 advantage against LightGBM for this problem.

\begin{table}[h!]
\centering
\setlength{\tabcolsep}{4pt}  
\begin{tabular}{l|r}
    Model & Relative AUC Gain vs. LightGBM \\
    \midrule
    MLP & -0.44 \% \\
    DCNv2 & -0.09 \% \\
    MLP + PLR & -0.23 \% \\
    LightGBM & (baseline) \\
    DCNv2 + PLR & +0.06 \% \\
    DCNv2 + PLR + L2 & +0.08 \% \\
\end{tabular}
\caption{DNN configurations vs.\ LightGBM baseline on tabular features.}
\label{tab:dnn_ablation_results}
\vspace{-20pt}
\end{table}

\section{Architecture and Training Details}
\label{sec:appendix}

\begin{table}[h]
    \centering
    \small
    \begin{tabular}{ll|ll}
        \hline
        \multicolumn{2}{c|}{\textit{DCNv2 Architecture}} & \multicolumn{2}{c}{\textit{Training}} \\
        \hline
        Cross network & Low-rank & Batch size & 2048 \\
        Deep network & [1024,512,256,128] & Epochs & 1 \\
        Tabular projection & 1024 & LR (transformer) & 1e-4 \\
        Numerical embed. & PLR, dim=8, freq=48 & LR (DCNv2) & 1.5e-4 \\
        Activation & GELU & Scheduler & Cosine \\
        \hline
        \multicolumn{2}{c|}{\textit{LoRA}} & \multicolumn{2}{c}{\textit{Optimizer}} \\
        \hline
        Rank & 64 & Adam $\beta_1, \beta_2$ & 0.9, 0.999 \\
        Dropout & 0.1 & Warmup ratio & 0.05 \\
        Target modules & c\_attn, c\_proj, c\_fc & Weight decay & 0.001 \\
        \hline
    \end{tabular}
    \caption{Model and training configuration.}
    \label{tab:config}
\end{table}

Table~\ref{tab:config} summarizes the complete model configuration. Joint hyperparameter search is computationally prohibitive (a single 203M training run requires 7 days on 8 A100s). We instead adopt a two-stage approach: (1) freeze the pre-trained transformer and search DCNv2 hyperparameters via 60 Ray Tune trials ($\sim$3 days on 5 H100s), then (2) perform end-to-end Joint Fusion with the selected configuration. This reduces search cost by $\sim$100$\times$.

\end{document}